\documentclass{aastex}
\usepackage{emulateapj5}
\lefthead{TSUJIMOTO \& SHIGEYAMA}
\righthead{Relics of subluminous supernovae}

\def\cm3{{\rm ~cm}^{-3}}
\def\ltsima{$\; \buildrel < \over \sim\;$}
\def\ltsim{\lower.5ex\hbox{\ltsima}}
\def\gtsima{$\; \buildrel > \over\sim \;$}
\def\gtsim{\lower.5ex\hbox{\gtsima}}
\def\ms{$M_{\odot}$ }
\def\msp{$M_{\odot}$}

\begin{document}
\title{Relics of subluminous supernovae in metal-poor stars}

\author{Takuji Tsujimoto$^{1}$ and Toshikazu Shigeyama$^{2}$}

\altaffiltext{1}{National Astronomical Observatory, Mitaka-shi,
Tokyo 181-8588, Japan; taku.tsujimoto@nao.ac.jp}

\altaffiltext{2}{Research Center for the Early Universe, Graduate
School of Science, University of Tokyo, Bunkyo-ku, Tokyo 113-0033,
Japan; shigeyama@resceu.s.u-tokyo.ac.jp}

\begin{abstract}
The unique elemental abundance pattern of the carbon-rich stars
CS29498-043 and CS22949-037 is characterized by a large excess of
magnesium and silicon in comparison with iron. This excess is
investigated in the context of a supernova-induced star formation
scenario, and it is concluded that these stars were born from the
matter swept up by supernova remnants containing little iron and that
such supernovae are similar to the least-luminous SNe ever observed,
SNe 1997D and 1999br. Comparison of the observed abundance pattern in
iron-group elements of subluminous supernovae with those of other
supernovae leads to an intriguing implication for explosion,
nucleosynthesis, and mixing in supernovae. The observed invariance of
these ratios can not be accounted for by a spherically symmetric
supernova model.
\end{abstract}

\section{INTRODUCTION}

Recently, \citet{Aoki_02} revealed a unique elemental abundance
pattern in a carbon-rich extremely metal-poor star CS29498-043. The
chemical composition of CS29498-043 has a large excess of
$\alpha$-elements such as Mg and Si as well as C and N compared to
Fe-group elements. A similar characteristic was previously identified
in another C-rich star CS22949-037 \citep{McWilliam_95, Norris_01,
Depagne_02}. In CS22949-037, oxygen is also found to be significantly
overabundant \citep{Depagne_02}. Since a mixing process at the
asymptotic giant branch of stellar evolution enhances exclusively C
and N, the lack of Fe is likely to give rise to the enhancements of C,
N, Mg and Si \citep{Norris2_01, Aoki_02}. The lack of Fe, of course,
must be initially imprinted in stars at formation because the low-mass
stars considered here are unable to destroy Fe nuclei in their
envelope.

Recent studies on the chemical compositions of metal-poor halo stars
have claimed that these stars might have inherited the abundance
pattern for the ejecta of the preceding few supernovae (SNe)
\citep{Audouze_95} or a single SN \citep[ST98 in the
following]{Shigeyama_98}. If this is the case, these stars would have
formed from the interstellar matter (ISM) comprising the ejecta of a
single SN. \citet{Depagne_02} compared the relative elemental
abundances observed in CS22949-037 with those of the theoretical SN
models proposed by \citet{Woosley_95} and obtained fairly good
agreement with a particular model. Furthermore, since the mass of the
ISM eventually swept up by the preceding SN is theoretically derived,
the observed abundance ratios with respect to H on the surfaces of
stars would allow the mass of each element contained in the SN to be
estimated \citep[ST98;][]{Tsujimoto_98}. A comparison of the derived
mass of each element with that of theoretical SN models also leads to
the conclusion that the apparent enhancement of $\alpha$-elements in
the above two stars is a result of the deficiency of Fe in the ejecta
of the preceding SN (see \S 2 for quantitative discussion). This
indicates that there existed subluminous SNe with a small amount of
$^{56}$Ni as a heat source, which decays to $^{56}$Fe through
$^{56}$Co with a half-life of $\sim$77 d.

\begin{figure*}[htb]
\begin{center}
\plotone{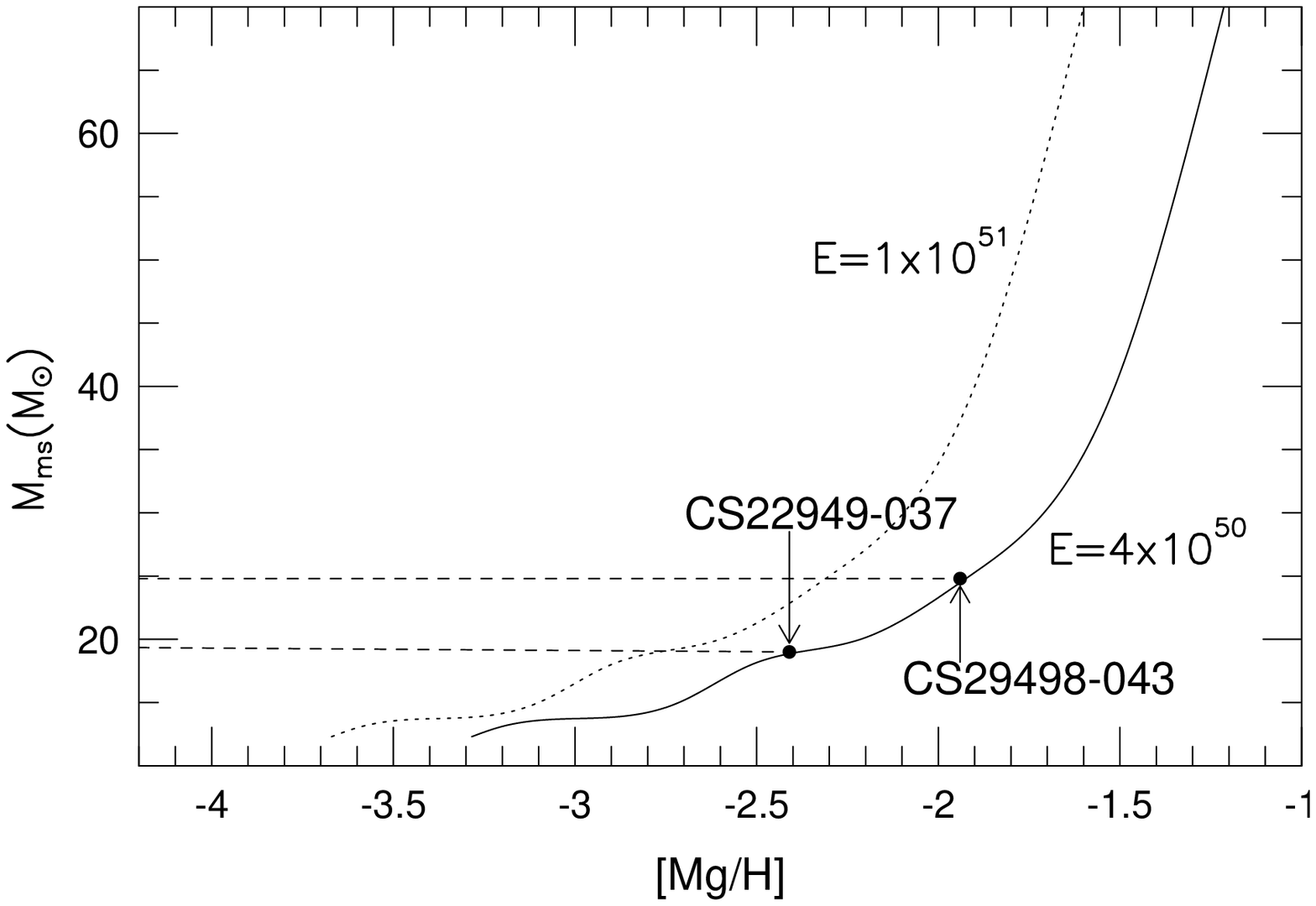}
\end{center}
\caption{SN II progenitor mass $M_{\rm ms}$ plotted as a function
of [Mg/H] inside the shell swept up by the SNR. Two cases for SN
explosion energy $E=4\times 10^{50}$ and $1\times 10^{51}$ erg are
shown. The observed [Mg/H] ratios for stars CS29498-043
\citep{Aoki_02} and CS22949-037 \citep{Depagne_02} are indicated by
{\it filled circles} on the curve for $E=4\times 10^{50}$ erg. Dashed
lines denote the SN progenitor masses corresponding to the
preceding SNe for these two stars.}
\end{figure*}

The least luminous SNe ever observed, SN 1997D \citep{Turatto_98} and
SN 1999br \citep[][and references therein]{Hamuy_02}, are in fact
Fe-poor. From analyses of the light curve of SN 1997D
\citep{Turatto_98, Benetti_01}, it has been found that this SN ejects
only $\sim 0.002\,M_\odot$ of $^{56}$Ni. The explosion energy $E$ is
estimated to be $E\sim 4\times 10^{50}$ erg and the progenitor mass to
be $\sim 25\, M_\odot$. \citet{Hamuy_02} reported that SN 1999br
ejects a similar amount of $^{56}$Ni ($\sim 0.0016 \,M_\odot$) with an
explosion energy of $\sim 6\times 10^{50}$ erg and that the ejected
mass, which gives the lower bound of the progenitor mass, has been
estimated to be 14 $M_\odot$ from the light curve analyses. It is well
known that there is a distinct class of SNe with similar progenitor
mass that yield Fe comparable to SN 1987A or slightly higher
($0.07-0.15 M_\odot$, e.g., ST98). Such SNe will be referred to as
{\it normal} SNe in the following. As mentioned in \citet{Turatto_98},
the low Fe mass of SN 1997D may be due to fallback of material onto
the collapsed remnant. In this case, Fe as well as other Fe-group
elements such as Cr, Mn, Co and Ni produced in the deep interior of
the core would fall back. In contrast, lighter elements synthesized in
the outer layers of the star would be ejected and be excluded from any
fallback. As a result, lighter elements are significantly enhanced
with respect to Fe-group elements. The observed elemental abundance
patterns of CS29498-043 and CS22949-037 exhibit characteristics that
fit this explanation.

In the next section, the features of SNe that induced the formation of
the stars CS29498-043 and CS22949-037 will be deduced from the
observed elemental abundance pattern in light of the SN-induced star
formation scenario proposed by ST98. Subsequently in \S 3,
nucleosynthesis for Fe-group elements is discussed based on the
implications of the [Cr, Mn, Co/Fe] ratios observed in these two stars,
which is surprisingly at variance with the present theoretical
expectations from spherically symmetric SN models.

\begin{figure*}[htb]
\begin{center}
\plotone{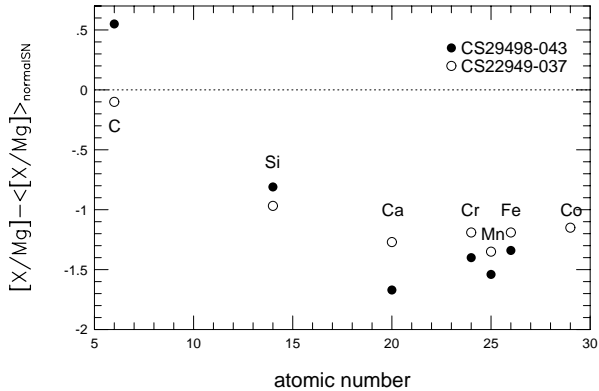}
\end{center}
\caption{Abundance ratios of elements with respect to Mg in
CS29498-043 ({\it filled circles}: Aoki et al.~2002) and CS22949-037
({\it open circles}: Depagne et al.~2002) relative to those in the
ejecta of normal SNe with the masses of 25 \ms and 20 \msp,
respectively. See text for details.}
\end{figure*}

\section{Stars with relics of subluminous supernovae}

The subluminous SNe discovered so far are less energetic varieties of
SNe, although theoretically some energetic SNe with high masses such
as 35 \ms could also be subluminous \citep{Woosley_95}. Such low energy
reduces the mass $M_{\rm SW}$ of ISM (mostly hydrogen) eventually
swept up by a subluminous SN, obeying the relation $M_{\rm SW} \propto
E^{0.97}$ (ST98). Since the metallicity [X/H] of extremely metal-poor
stars born from the ISM swept up by a SN remnant (SNR) is inversely
proportional to $M_{\rm SW}$, stars that inherit the ejecta of a
subluminous SN should have larger [X/H] as long as the other
conditions are the same.

As discussed in ST98, the abundance ratios [Mg/H] in metal-poor stars
can be used as an indicator of the progenitor mass of the SN inducing
the formation of the stars, since theoretical models for massive stars
predict that the mass of Mg inside a star increases with increasing
stellar mass \citep[e.g.,][]{Woosley_95, Umeda_02}. In addition,
the amount of ejected lighter elements such as Mg can be predicted
more precisely than that of heavier elements such as Fe because the
ejection of the outer layer is less affected by theoretical
uncertainty concerning the SN explosion mechanism \citep[see, however,
model Z20A of][ for example]{Woosley_95}.

The relationships between the metallicity [Mg/H] of stars and the mass
$M_{\rm ms}$ of the SN progenitor thus obtained are shown in Figure 1
for $E=4\times 10^{50}$ erg ({\it solid line}) and $E=1\times10^{51}$
erg ({\it dotted line}), assuming the metallicity [Mg/H] of stars born
from the shell formed by an SNR is approximated well by the average
[Mg/H] inside the ISM swept up by the SNR. If CS29498-043 inherited
the ejecta of a subluminous SN with explosion energy equal to that of
SN 1997D, the SN progenitor mass becomes 25 \msp, which intriguingly
agrees with that inferred from the light curve analyses of SN
1997D. The SN progenitor mass for CS22949-037 estimated in the same
manner is $M_{\rm ms}\approx20$\msp. The masses of ejected Fe are
estimated to be 0.006 \ms (CS29498-043) and 0.003 \ms (CS22949-037)
using the observed [Fe/H] ratios for the two stars and assuming
$E=4\times 10^{50}$ erg. Both of the estimated Fe masses indicate
subluminous SNe such as SNe 1997D and 1999br.

Figure 2 shows the enhancement factors of each element defined by the
abundance ratios of the element with respect to Fe in the metal-poor
stars discussed in this letter relative to those ratios in the
ejecta of normal SNe with the corresponding progenitor masses. The
abundances of normal SNe have been inferred from the elemental
abundance pattern observed for metal-poor stars in the manner
described in \citet{Tsujimoto_98}. Thus derived abundance ratios
correspond to the average values of metal-poor stars observed by
\citet{McWilliam_95} at the metallicity of [Mg/H]$\approx -2.3$ and
$-2.8$, respectively (see Fig.~1). That is, an SN with
$E=1\times10^{51}$ erg and $M_{\rm ms}=25$ \ms will imprint a
metallicity [Mg/H]$=-2.3$ on the descendant stars, while an SN with
the same energy but with $M_{\rm ms}=20$ \ms corresponds to
[Mg/H]$=-2.8$.  As seen in Figure 2, the abundance of elements heavier
than Si in these stars is at least a factor of 5 smaller than the
average value of metal-poor stars, which is attributed to fallback
onto the remnant in subluminous SNe. A moderate enhancement factor for
Si in comparison with Fe-group elements might imply that fallback
occurred inside the Si layer. The enhancement factors for Ca through
Co have similar values in CS22949-037, while Cr through Fe are simlar
in CS29498-043. A good coincidence in the enhancement factors is seen
especially in CS22949-037, probably due to more accurate abundance
determinations. This result suggests that there is little difference
in the ratios of Ca, Cr, Mn, or Co to Fe between subluminous SNe and
normal SNe with the same progenitor mass, with intriguing implications
for nucleosynthesis, explosion, and mixing in SN models.

\section{Implications for explosive nucleosynthesis and supernova
explosion}

Figure 3 shows the observed correlations between [Cr/Fe], [Mn/Fe],
[Co/Fe] and [Mg/H] for metal-poor halo stars \citep{McWilliam_95},
CS29498-043, and CS22949-037. As shown in these figures, the observed
[Cr, Mn, Co/Fe] trends are quite clear; the [Cr, Mn/Fe] ratios
increase with increasing [Mg/H], whereas the [Co/Fe] ratio decreases.
These clear correlations suggest that the [Cr/Fe], [Mn/Fe], [Co/Fe]
ratios are closely correlated with the SN progenitor mass
\citep{Tsujimoto_98, Nakamura_99}, while the abundance ratios of
$\alpha$-elements such as Mg, Si, and Ca to Fe do not have clear
correlations. It is also clear from the same figure that the observed
[Cr/Fe], [Mn/Fe], [Co/Fe] ratios of CS29498-043 and CS22949-037 do not
follow the trend presented by the other stars.  Five horizontal arrows
in this figure indicate how these ratios would shift when these stars
inherited the ejecta of an SN with $E=1\times10^{51}$ erg instead of
$E=4\times10^{50}$ erg. The lengths of the horizontal shifts
correspond to differences in the swept-up hydrogen mass between SNe
with $E=4\times 10^{50}$ and $1\times10^{51}$ erg.  All the arrows
pointing to the lines of the best fit for the observational data by
$\chi^2$ test, suggesting that the [Mn/Fe], [Cr/Fe], and [Co/Fe]
ratios of these two stars do not deviate from those of normal SNe with
the same progenitor mass, although for CS29498-043 an uncertainty
remains due to the large error in the observed [Mn/Fe] ratio and a
lack of Co data. As a consequence, there is little difference in the
ratios of Cr, Mn, or Co to Fe between subluminous SNe and normal SNe
given the same progenitor mass. This turns out to be incompatible with
spherically symmetric SNe, as follows.

The elements Mn and Cr are the products of incomplete Si burning,
whereas Co is produced by complete Si burning. Fe is produced in both
burning processes. The region of complete Si burning is located at the
bottom of the ejecta. If an SN explosion were spherically symmetric,
different amounts of Fe could only be attributable to different
combinations of the mass cut separating the ejecta from the remnant
and the explosion energy. Since the mass cut determines the ejected
fraction of the region in which complete Si burning takes place,
smaller Co/Fe ratios would be expected from SNe with smaller amounts
of Fe, corresponding to subluminous SNe. Larger Mn/Fe and Cr/Fe ratios
would be expected from these SNe, as shown by \citet{Umeda_02}. Lower
explosion energy would suppress complete Si burning and thus reduce
both the amount of Fe and the Co/Fe ratio. Therefore, any combination
of explosion energy and mass cut in spherically symmetric SNe is
incompatible with the observed invariance of the abundance ratios of
Fe-group elements to Fe. Since Rayleigh-Taylor instabilities take
place after the nuclear reactions freeze out as was shown by a
multi-dimensional calculation \citep{Kifonidis_00}, mixing does not
alter the relative abundances of Fe-group elements determined in the
way mentioned above. The same calculation has also shown that the
mixing process is effective. Thus it is expected that fallback that
may occur afterwards would hardly change these relative abundances,
either. Mixing is unlikely to explain the observed invariance of the
relative abundances among SNe with different explosion energies. 

We will discuss a possibility that the observed ratios among Fe-group
elements for the two stars reflect those in the ISM swept up by an SNR
rather than those in individual SN ejecta.  Indeed, the gas swept up
by the SNR from which stars are formed contains not only heavy
elements ejected from the SN itself but also those already present in
the ISM. In general, for metal-poor regime such as [Mg/H] \ltsim --2,
the stellar abundance pattern is essentially determined by the SN
yield due to a negligible contribution from the ISM
\citep[ST98;][]{Tsujimoto_99}. A lack of iron-group elements in the
ejecta of a sub-luminous SN could enhance contribution from the ISM to
the stellar abundance of iron-group elements. Since the [Cr/Fe],
[Mn/Fe], [Co/Fe] ratios in the ISM should be equal to nearly zero,
which corresponds to the plateau value seen in halo stars with
--2$<$[Fe/H]$<$--1, the two stars discussed in this letter should show
these abundance ratios nearly equal to the solar values if this is the
case. However, the observed ratios shown in Figure 3 do not support
this possibility. Therefore it is likely that these stars contain
little heavy elements from the ISM.  Furthermore, combinations of
these ratios predicted from spherically symmetric SN models with those
in the ISM can not make the observed ones.

\begin{figure*}[htb]
\begin{center}
\plotone{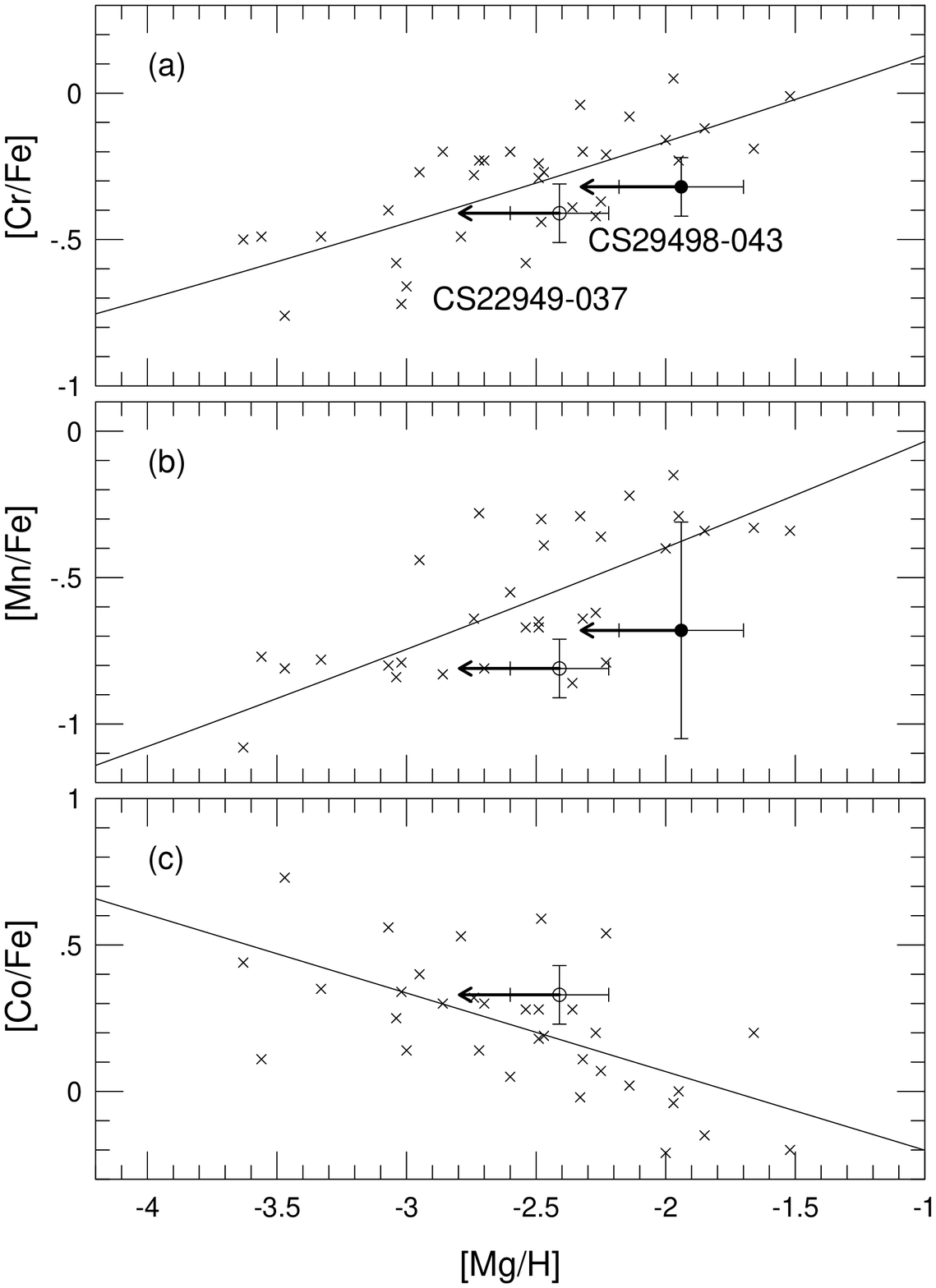}
\end{center}
\caption{
($a$) The correlation between [Cr/Fe] and [Mg/H] for
metal-poor stars observed by \citet{McWilliam_95}, CS299498-043
\citep{Aoki_02} and CS22949-037 \citep{Depagne_02}.  The solid line
shows the $\chi^2$ fit to the observed points. Arrows indicate how the
[Mg/H] ratios for these stars would shift if these stars were born
from the SNRs originating from SNe with $E=1\times 10^{51}$
ergs. ($b$) The same as ($a$) but for Mn. ($c$) The same as ($a$) but
for Co.}
\end{figure*}

Stellar roation might be a factor to produce a variety in energies of
supernova explosions with the same progenitor mass. It is sometimes
suggested that a rapid rotaion leads to a jet-like explosion
\citep[e.g.,][]{Woosley_93}. In fact, it has been a preferred
explanation for matter mixing suggested by the emission line profiles
observed for SN 1987A \citep{Nagataki_00} and SN 1998bw
\citep{Maeda_02}, both of which are core-collapse SNe. Furthermore,
the optical spectropolarimetry of several core-collapse SNe suggests
highly asymmetric explosion mechanisms such as the formation of
bipolar jets \citep{Wang_01}. Detailed numerical studies on
nucleosynthesis and ejection mechanism in jet-like explosions might be
required to explain the observed trend of Fe-group elements in halo
field stars.

An alternative solution may be in a hot bubble formed by neutrino
emission. If neutrinos from the nascent neutron star deposit
substantial energy to make a hot bubble behind the blast wave, the hot
bubble may carry exclusively the product of complete Si burning to the
outer layer by buoyancy. A numerical simulation by \citet{Wilson_88}
has shown that this mechanism leads to a less energetic SN. This may
suppress fallback of Co and Fe in subluminous SNe, though other
numerical studies have shown that a hot bubble does not result in a SN
explosion \citep[e.g.,][]{Bruenn_95}. The current understanding of
the involved physics is, however, still too premature to extract a
definite conclusion on this issue.

\section{CONCLUSIONS}

The recently discovered new class of carbon-rich stars including
CS29498-043 and CS22949-037, which exhibit a significant enhancement
of $\alpha$-elements with respect to Fe-group elements, has been shown
in this study to inherit the elemental abundance pattern of a
precursor SNe that ejected a small but adequate amount of Fe with low
explosion energy. These SNe are found to resemble the two recently
discovered subluminous SNe 1997D and 1999br in terms of the ejected Fe
mass, explosion energy, and progenitor mass. Subluminous supernovae
may therefore be studied more easily by looking at the surface of
nearby metal-poor stars than by searching for similar occurrences
elsewhere.

The present study revealed that there is little difference in the
[Cr/Fe], [Mn/Fe], [Co/Fe] ratios between subluminous SNe and normal
SNe given the same progenitor mass, which has intriguing implications
in nucleosynthesis and explosion in SN models. The observed facts are
apparently at odds with the prediction by spherically symmetric SN
explosion models.

\acknowledgements

We are grateful to the anonymous referee for making useful comments.
This work was supported in part by a Grant-in-Aid for Scientific
Research (11640229, 12640242) from the Ministry of Education, Culture,
Sports, Science, and Technology of Japan.

\end{document}